\documentclass[twocolumn,amssymb,fleqn]{revtex4-1} %fleqn: make all eqns left alighed  ,showpacs
\usepackage{epsfig,amssymb,amsmath,graphicx,subfigure,hyperref  }  % pdfpages and graphicx are added   hyperref(set hyperlink for content)
\usepackage{color,bm} \usepackage[normalem]{ulem}% The line notes delete
\usepackage{multirow} \usepackage{tabularx}

\usepackage{color,bm}
\usepackage[normalem]{ulem}

\begin{document}
\title{Geometry and physics in the deformations of
crystalline caps} \author{Jingyuan Chen and Zhenwei Yao} 
\email{zyao@sjtu.edu.cn}
\affiliation{School of Physics and Astronomy, and Institute of Natural
Sciences, Shanghai Jiao Tong University, Shanghai 200240, China}
\begin{abstract} 
  Elucidating the interplay of stress and geometry is a fundamental
  scientific question arising in multiple fields. In this work, we
  investigate the geometric frustration of crystalline caps confined on the
  sphere in both elastic and plastic regimes. Based on the revealed
  quasi-conformal ordering, we discover the partial, but uniform screening of
  the substrate curvature by the induced curvature underlying the inhomogeneous
  lattice. This scenario is fundamentally different from the conventional
  screening mechanism based on topological defects. In the plastic regime, the
  yield of highly stressed caps leads to fractures with featured morphologies
  not found in planar systems. We also demonstrate the strategy of engineering
  stress and fractures by vacancies. These results advance our general
  understanding on the organization and adaptivity of geometrically-frustrated
  crystalline order.
\end{abstract}

\maketitle
%Begin of Main Text
\section{Introduction}
Condensed matters confined on curved geometries are widely seen in multiple
fields,~\cite{nelson2002defects, bowick2009two} ranging from protein
shells~\cite{caspar1962physical, lidmar2003virus, zandi2004origin} to liquid
interfaces coated by colloids~\cite{dinsmore2002colloidosomes, bausch2003grain,
	ershov2013capillarity} and liquid-crystals.~\cite{nelson2002toward,
	fernandez-nieves2007novel} The intrinsic conflict of two-dimensional ordering
and substrate curvature creates unique stress patterns.~\cite{
	grason2013universal, azadi2014emergent} Especially, resolving the accumulated
stress in regular particle arrays leads to rich physics associated with the
elastic~\cite{grason2013universal, vernizzi2011platonic} and
plastic~\cite{azadi2014emergent, li2019ground, yao2017topological,
	mitchell2017fracture} response of the system, which also underlies a host
of problems in materials geometry. Examples include the curvature driven
patterned deformations~\cite{grason2013universal, vernizzi2011platonic,
	klein2007shaping} and the exceedingly rich topological defect
structures.~\cite{bausch2003grain, bowick2002crystalline, irvine2010pleats,
	kusumaatmaja2013defect} While much has been learned about the elastic patterns
and defect motifs in curved 2D crystals,~\cite{nelson2002defects, bowick2009two,
	audoly2010elasticity} the microscopic process of the transition from the elastic
deformation to the yield of the crystalline order and its connection with stress
has not yet been fully explored.

The goal of this work is to elucidate these fundamental questions regarding the
adaptivity of crystalline order on curved space.~\cite{li2019ground,
	yao2017topological} The spherical crystalline cap provides a suitable model to
address these questions.~\cite{grason2013universal, li2019ground,
	yao2017topological} The crystalline cap is composed of a regular array of point
particles that interact by the Lennard-Jones (L-J)
potential.~\cite{jones1924determination} In this work, by applying the fixed
boundary condition, the stretching of the cap could be precisely controlled by
shrinking the sphere.  By the combination of geometric analysis and elasticity
theory, we analytically analyze the inhomogeneous
packings of the particles in mechanical equilibrium, reveal the quasi-conformal ordering underlying the
inhomogeneous lattice, and discover the partial, but uniform screening of the
substrate curvature by the inhomogeneity-induced curvature. This scenario is
fundamentally different from the conventional screening mechanism based on
topological defects.~\cite{Nelson1987,bowick2000interacting} 
The highly stressed cap ultimately experiences plastic
deformation. The featured morphology of the fracture and the associated
energetics are discussed.  We also demonstrate the strategy of exploiting the
stress-concentration effect of vacancies~\cite{timoshenko1951theory,
	landau1959theory} to engineer stress and fractures. This work demonstrates the
rich physics in the deformations of crystalline caps, and may have implications
in the engineering of extensive crystalline materials.

\section{Model and method}
\begin{figure*}[!htb] 
	\subfigure[]{ \includegraphics[width=0.17\textwidth]{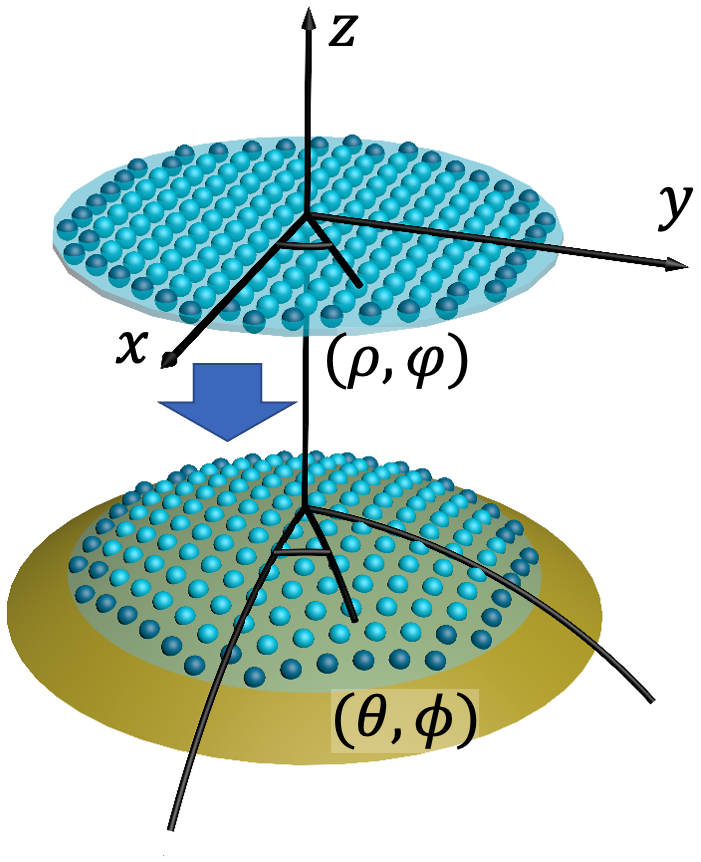}} \hspace{0.0in} \vspace{0.1in}
	\subfigure[]{ \includegraphics[width=0.26\textwidth]{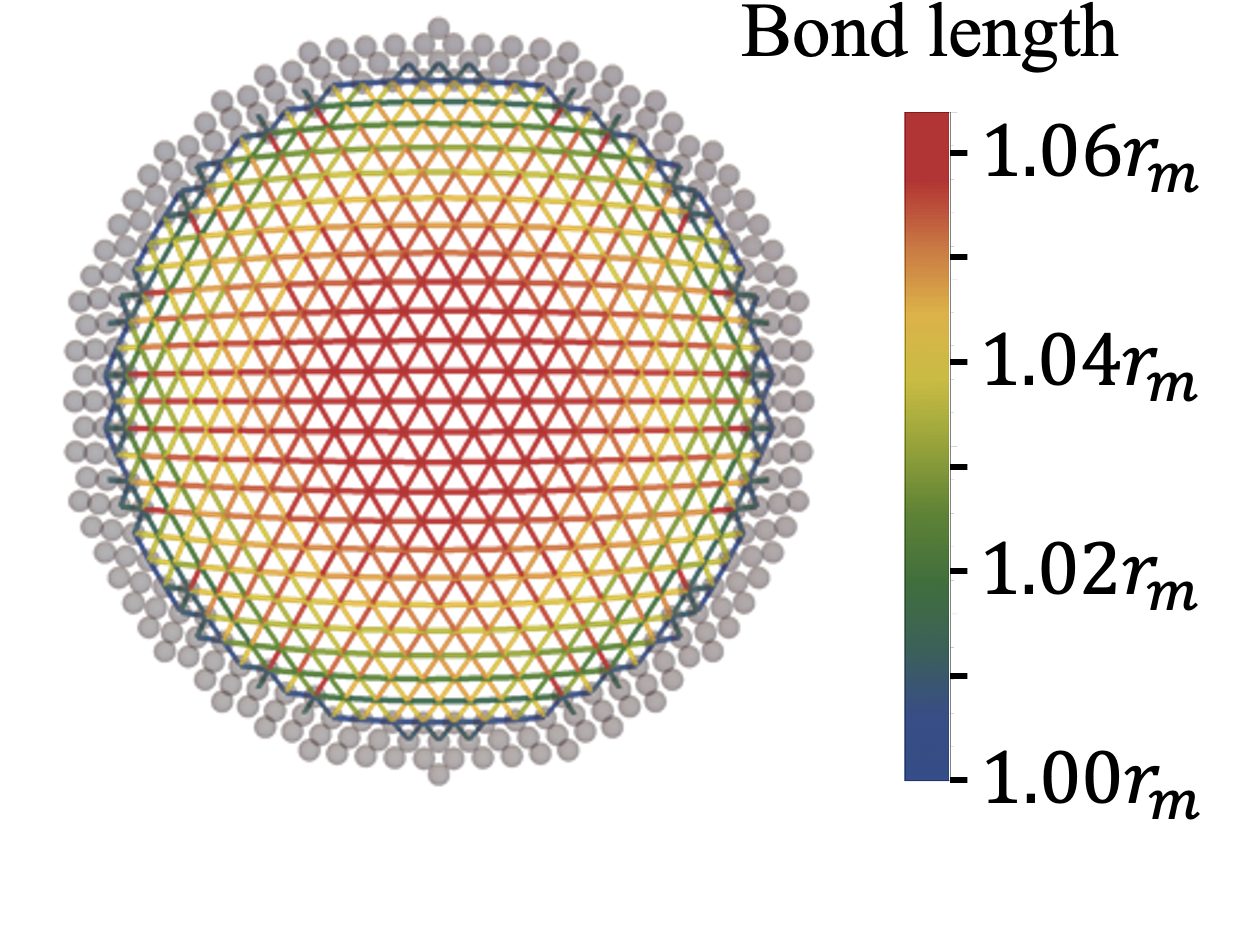}}
	\subfigure[]{ \includegraphics[width=0.26\textwidth]{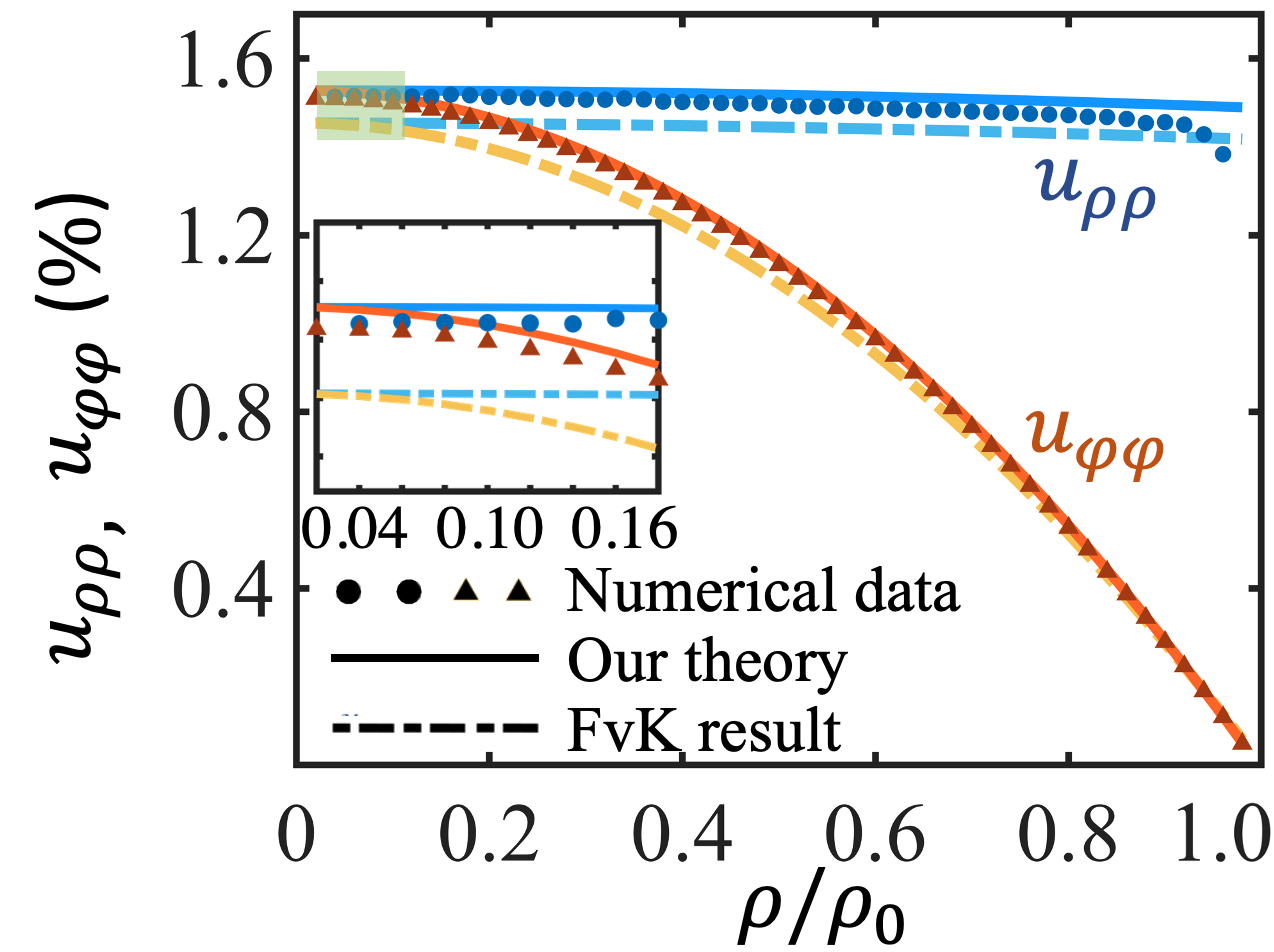}}
	\subfigure[]{ \includegraphics[width=0.26\textwidth]{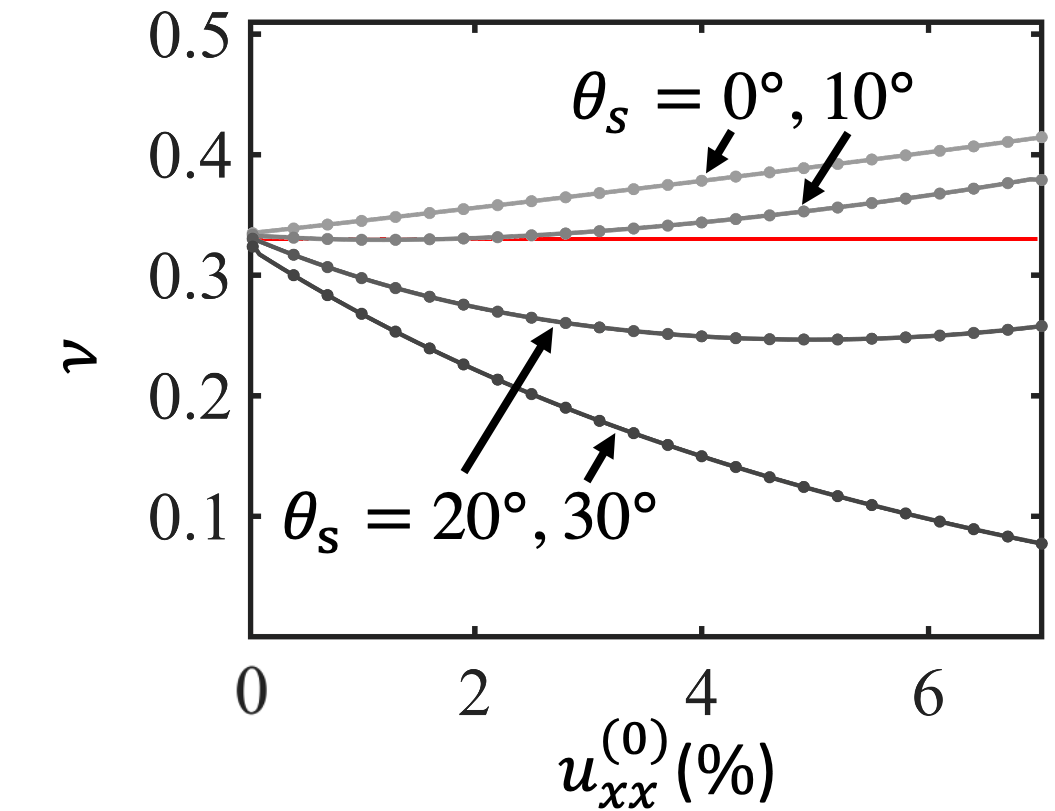}}
	\caption{Elastic deformation of the crystalline cap confined on the
		sphere. (a) A schematic plot of the crystalline cap model. The darker dots
		represent the anchored boundary particles. Polar coordinates $(\rho,\varphi)$ and spherical coordinates 
		$(\theta,\phi)$ are established over the undeformed planar disk and the
		deformed cap, respectively. Note that $\varphi = \phi$ due to the
		rotational symmetry in the deformation. 
		(b) The top view of an inhomogeneously stretched L-J crystalline cap in mechanical equilibrium.
		$\gamma=5.95\%$. $\rho_0=10r_m$. $R=18.1r_m$.  (c) The strain field in the
		stretched cap. $\gamma=1.5\%$. $\rho_0=50r_m$. $R=170.1r_m$.
		$\nu_{\text{eff}}= 0.31$. (d) Dependence of the Poisson's ratio on the
		preset strain $u^{(0)}_{xx}$ of a planar triangular L-J lattice. $\theta_s$ is
		the angle of the direction of stretching and the principal axis of the lattice
		(the x-axis). The red line at $\nu=1/3$ indicates the reference value
		of the Poisson's ratio for the isotropic elastic medium composed of linear 
		springs in triangular lattice. }\label{Fig1} 
\end{figure*}
In our model, the L-J potential allows us to conveniently explore both the
elastic and plastic regimes; the formally simple L-J potential has also been
extensively used to model various chemical and physical bonds
.~\cite{israelachvili2011intermolecular} $V_{\scriptscriptstyle
	\text{L-J}}(r)=\varepsilon_0[(r_m/r)^{12}-2(r_m/r)^6]$, where $r$ is the
Euclidean distance between two particles, $r_m$ is the equilibrium distance, and
$-\varepsilon_0$ is the minimum potential energy. All of the particles are
confined on the sphere of radius $R$ by geometric constraint without resorting
to any external potential.  The particles on the boundary annulus of the cap are
anchored, which constitutes the fixed boundary condition [see
Fig.~\ref{Fig1}(a)]. The cap is stress free in the limit of infinitely large
$R$. The degree of the stretching is characterized by $\gamma = (\omega_{0}-
\rho_0)/\rho_0$, where $\rho_0$ is the radius of the circular boundary and
$\omega_{0}$ is the geodesic radius of the spherical cap.  $\omega_{0}=R \arcsin
(\rho_0/R) \approx \rho_0 + K_s\rho_0^3/6$, where $K_s$ is the
Gaussian curvature of the spherical substrate of radius
$R$.~\cite{books/daglib/0090942}  $\omega_{0}= \rho_0$ as $R\rightarrow \infty$.
In simulations, the radius of the sphere is gradually reduced. For each given
$R$, the system is fully relaxed to the lowest energy state by the movement of
the particles on the surface of the sphere according to the high-precision
steepest descent method.~\cite{snyman2005practical, yao2016electrostatics, SI}
In this work, the units of length and energy are $r_m$ and $\varepsilon_0$.

\section{Results and discussion}

We first analytically analyze the elastic deformation of the crystalline cap by
the continuum elasticity theory. The packing of the L-J
particles in mechanical equilibrium is determined by the force balance equations
in terms of the stress tensor $\bm{\sigma}$.~\cite{landau1959theory}
In spherical coordinates ($\theta$, $\phi$), the in-plane force balance equation of $\nabla \cdot
\bm{\sigma}=0$ becomes 
\begin{align}
&\partial_{\theta}\sigma_{\theta\theta}+\partial_{\phi}\left(\frac{\sigma_{\theta\phi}}{\sin\theta}\right)+\cot\theta\left(\sigma_{\theta\theta}-\sigma_{\phi\phi}\right)=0,\label{stress_sphere2a}\\
&\partial_{\theta}\sigma_{\theta\phi}+\frac{1}{\sin\theta}\partial_\phi\sigma_{\phi\phi}+2\cot\theta\sigma_{\theta\phi}=0.
\end{align}
To deal with the boundary condition, we express the force balance equations in
terms of the displacement $u_{\rho}$ over the undeformed planar disk, where the
polar coordinates $(\rho, \varphi)$ are established [see
Fig.~\ref{Fig1}(a)]. The detailed derivation is presented in Supplemental
Material.~\cite{belytschko2013nonlinear,SI} By applying the boundary conditions
of $u_{\rho}(0)=u_{\rho}(\rho_0) =0$ and making use of the rotational symmetry
of the system, we obtain the analytical expressions for the strain field up to
$(\rho_0/R)^4$: 
\begin{align}
\label{Solutionu11} 
\begin{split}
u_{\rho\rho}=&\frac{3-\nu}{16}\left(\frac{\rho_0}{R}\right)^2+\left[\frac{7\left(3-\nu\right)^2}{512}\left(\frac{\rho_0}{R}\right)^4 \right]\\
&+\left \{ \frac{3\nu-1}{16}\left(\frac{\rho_0}{R}\right)^2 \right. \\ 
&\ \ \ \ \ \  \left.+\left[\frac{(3\nu-1)(3-\nu)}{64}\left(\frac{\rho_0}{R}\right)^4\right]\right\} \frac{\rho^2}{\rho_0^2},
\end{split}
\end{align}
\begin{align}
\label{Solutionu22} 
\begin{split}
u_{\varphi\varphi}=&\frac{3-\nu}{16} \left(\frac{\rho_0}{R}\right)^2 + \left[\frac{7\left(3-\nu\right)^2}{512}\left(\frac{\rho_0}{R}\right)^4 \right]\\
&+\left\{\frac{\nu-3}{16}\left(\frac{\rho_0}{R}\right)^2 \right.\\ 
&\ \ \ \ \ \ \left.+\left[\frac{-(3-\nu)^2}{64}\left(\frac{\rho_0}{R}\right)^4 \right]\right\}
\frac{\rho^2}{\rho_0^2}.
\end{split} 
\end{align}
The shear
components of the strain tensor are zero.  In Eqs.(\ref{Solutionu11}) and
(\ref{Solutionu22}), the terms in the square brackets are the higher order correction
terms in comparison with the conventional approximate solutions to the FvK
equations. The FvK equations describe the out-of-plane deformation of
plates.~\cite{landau1959theory}. The axisymmetric form of the FvK equations
is:\cite{grason2013universal, king2012elastic}
\begin{align}
\label{FvK_radial}
&\frac{d}{d\rho}(\rho\sigma_{\rho\rho})-\sigma_{\varphi\varphi} = 0,\\
\label{FvK_normal}
&B\Delta^2\zeta-\sigma_{\rho\rho}\frac{d^2\zeta}{d \rho^2}-\frac{\sigma_{\varphi\varphi}}{\rho}\frac{d \zeta}{d \rho}=F_N,
\end{align}
$\sigma_{\rho\rho}$ and $\sigma_{\varphi\varphi}$ are the in-plane stress
tensors in polar coordinates on the undeformed plate. $\zeta(\rho)$ is the
out-of-plane deformation. $B$ is the bending modulus, $F_N$ is the exerted
normal force per unit area and $\Delta$ is the Laplacian operator.  By
expressing Eq.~(\ref{FvK_radial}) in terms of $u_\rho$ and applying the boundary
conditions $u_\rho(0)=u_\rho(\rho_0)=0$, we obtain the expressions for the
strain field up to the order of $(\rho_0/R)^2$, as contained in Eqs.
(\ref{Solutionu11}) and (\ref{Solutionu22}).~\cite{SI}

Equations (\ref{Solutionu11}) and (\ref{Solutionu22}) show that the homogeneously curved 
substrate geometry creates an inhomogeneous strain field, which is fundamentally
different from the case of a planar membrane under radial tension. Simulations
confirm the inhomogeneous distribution of the L-J particles on the sphere, as
shown in Fig.~\ref{Fig1}(b).  Analysis of the bond length, which are constructed
by the standard Delaunay triangulation procedure,~\cite{nelson2002defects} shows
that the central region in the deformed lattice is subject to a stronger
stretching. Furthermore, the ratio of the maximum to the minimum bond length in
the equilibrium configuration increases with the value of $\gamma$, indicating
that the degree of inhomogeneity is enhanced as the sphere shrinks.

In Eqs.~(\ref{Solutionu11}) and (\ref{Solutionu22}), we notice that
the strain field is completely determined by the Gaussian curvature $1/R^2$ of
the substrate and the Poisson's ratio $\nu$; the Young's modulus does not enter the
equation. It is known that $\nu=1/3$ for an isotropic elastic medium composed of
linear springs in triangular lattice.~\cite{landau1959theory, seung1988defects}
However, since the residual strain could influence the value of $\nu$,
~\cite{greaves2011poisson} could we still specify a uniform value
to the Poisson's ratio for our system?

\begin{figure}[b!] \subfigure[]{
		\hspace{-0.19in}
		\includegraphics[width=0.248\textwidth]{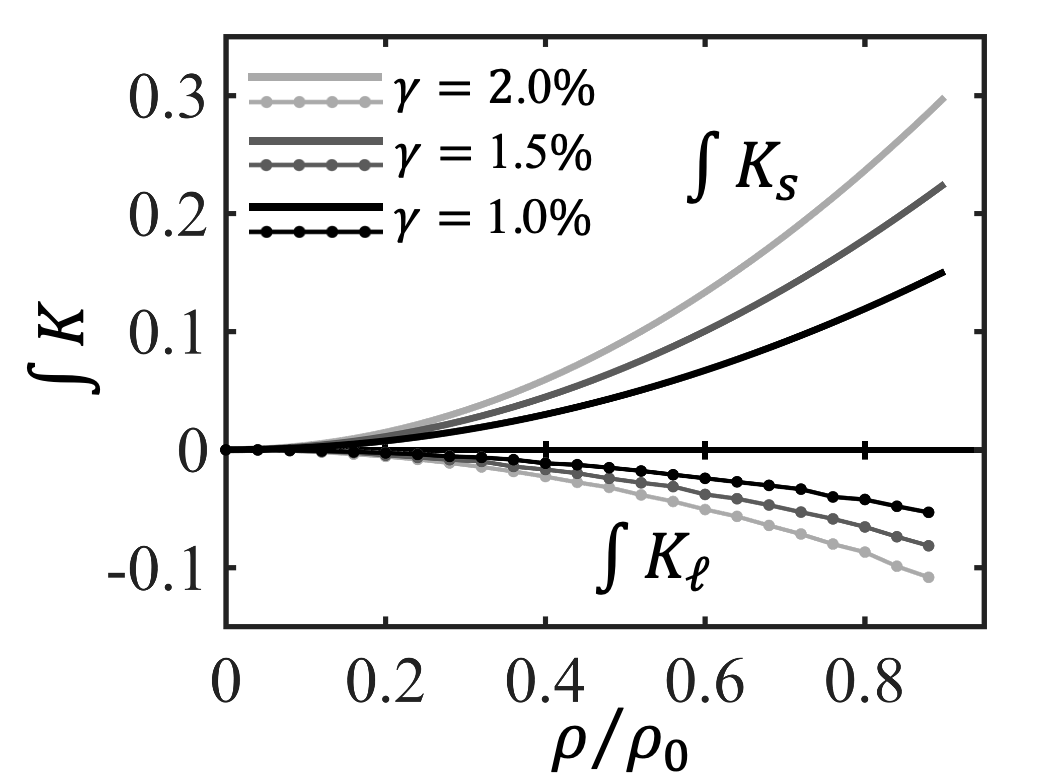}}
	\hspace{-0.1in}
	\subfigure[]{
		\includegraphics[width=0.248\textwidth]{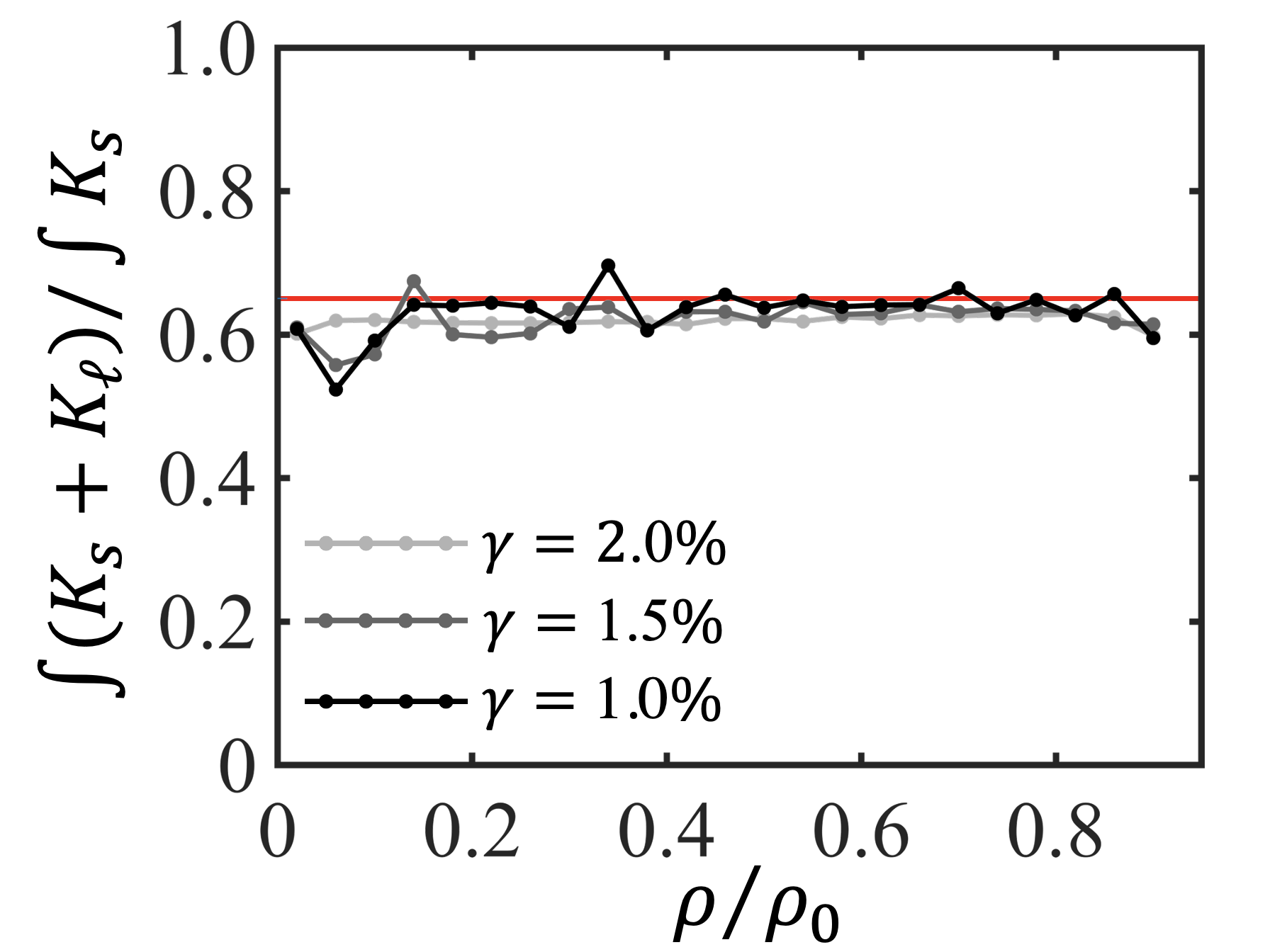}}
	\hspace{-0.15in}
	\caption{Numerical simulations reveal the partial, but uniform 
		screening of the substrate curvature $K_s$
		by the induced curvature $K_\ell$ underlying the inhomogeneous packings of
		the L-J particles. (a) Plot of the integrated $K_s$ and $K_\ell$ at varying
		degrees of stretching. (b) Plot of the ratio of the excess integrated
		curvature and the integrated $K_{s}$. The stable curves indicate the uniform
		screening of the substrate curvature by $K_\ell$. The red line 
		indicates the theoretical value of $0.65$.  $\rho_0=50r_m$. } \label{Fig2} 
\end{figure}

\begin{figure*}[!ht] 
	\subfigure[]{ 
		\includegraphics[width=0.43\textwidth]{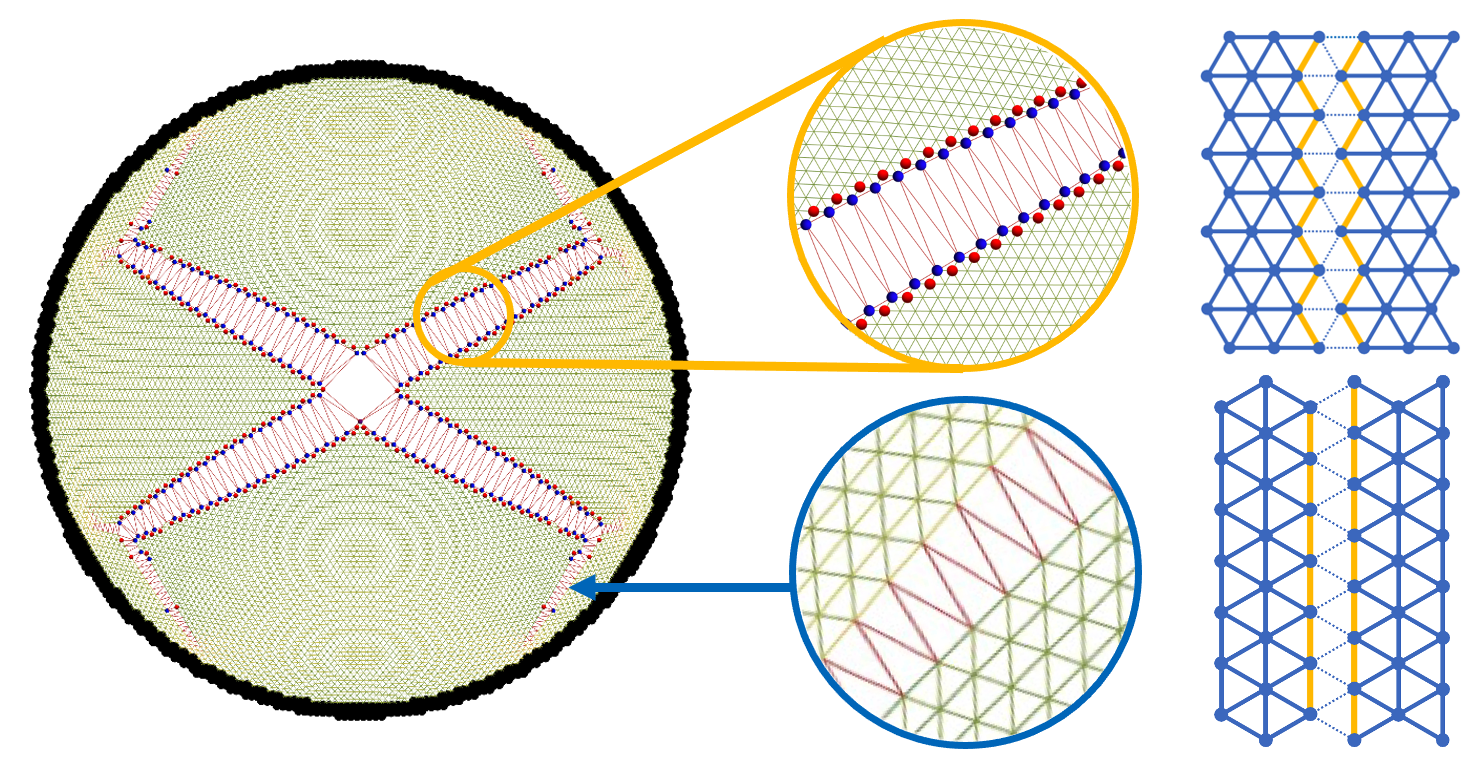}} 
	\subfigure[]{ 
		\includegraphics[width=0.25\textwidth]{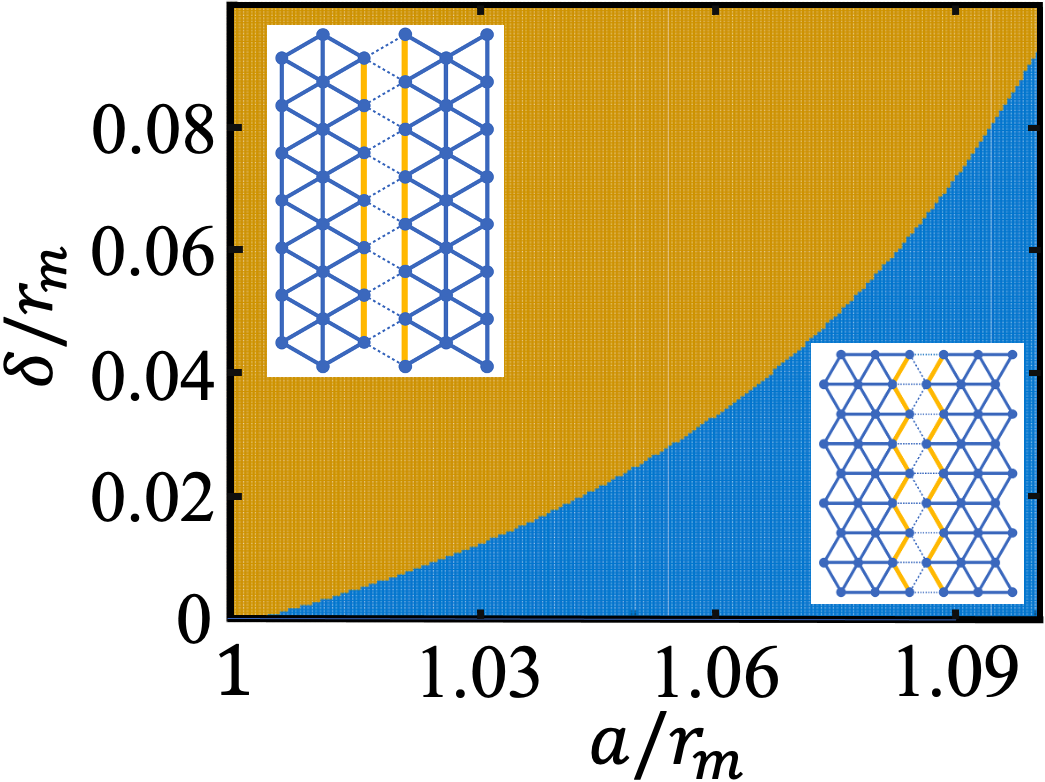}} 
	\hspace{0.1in} 
	\subfigure[]{
		\includegraphics[width=0.25\textwidth]{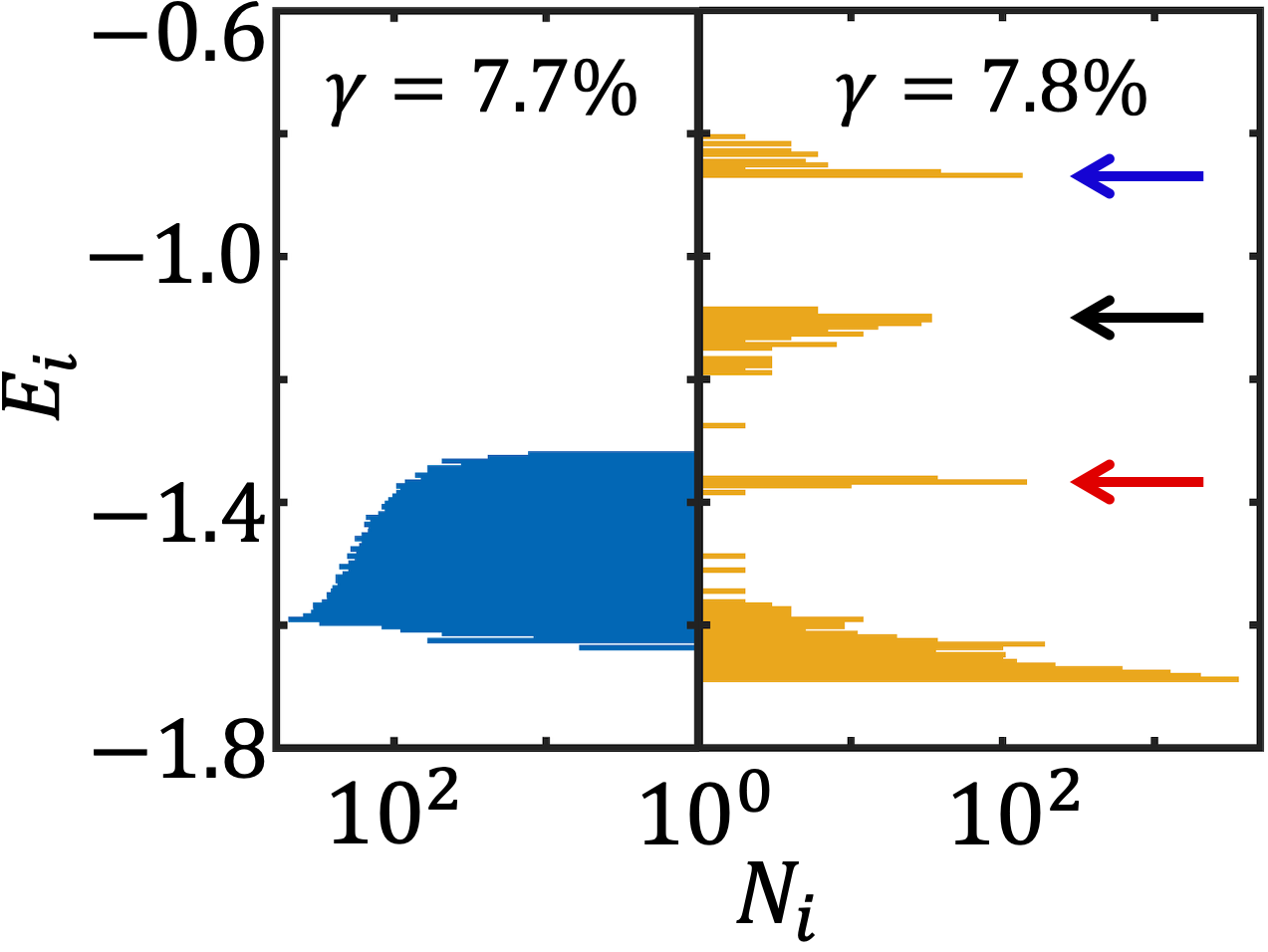}} 
	\caption{The yield of the crystalline cap with the shrinking of the sphere.
		(a) A typical fracture pattern featured with a $90^{o}$ turn at the
		tip. The insets show the zigzag and flat fractures. The red and blue
		dots represent five- and seven-fold disclinations. $\gamma=7.8\%$. $\rho_0=50r_m$. 
		(b) The phase diagram for the emergence of zigzag and flat fractures in the
		isotropically stretched planar L-J lattice of lattice spacing $a$. 
		$\delta$ is the relative displacement of the crystals separated by the yellow boundaries. (c) The
		redistribution of the energy of individual particles caused by the rupture
		event. The single-peak profile prior to rupture ($\gamma=7.7\%$) splits into
		multiple energy bands at $\gamma=7.8\%$. $N_i$ is the number of particles
		whose energy is in the range of $E_i\pm \delta E/2$. $\delta E=0.015$. 
		The elevated energy bands indicated by the blue and red arrows 
		correspond to the particles of the same color along the zigzag fractures; 
		see the upper circle in (a). And the energy band indicated by the black 
		arrow corresponds to the particles along the flat fracture; see the lower 
		circle in (a). $\rho_0=50r_m$.}
	\label{Fig3} \end{figure*} 

To address this question, we numerically study the dependence of the Poisson's
ratio on the magnitude of the preset strain and the direction of stretching.
The detailed information about the measurement of Poisson's ratio is presented
in Supplemental Material.~\cite{SI} From Fig.~1(d), we see that the nonmonotonous
$\nu$-curves converge to a common value that is very close to $1/3$ as
$u^{(0)}_{xx}$ becomes vanishingly small.  It suggests that the Poisson's ratio
of a slightly stretched crystalline cap may
be approximately specified by a uniform value. As such, the parameter $\nu$
appearing in Eqs.(\ref{Solutionu11}) and (\ref{Solutionu22}) shall be
interpreted as an effective Poisson's ratio,~\cite{greaves2011poisson} whose value is determined as
$\nu_{\text{eff}}=0.31$ by matching the analytical result and the simulation
data for $\gamma=1.5\%$.  $\nu_{\text{eff}}=0.31\pm0.01$ for $\gamma \in [1\%,
2\%]$.

We plot the theoretically obtained strain field in Fig.~1(c) by inserting the
value of $\nu_{\text{eff}}$ into Eqs.~(\ref{Solutionu11}) and
(\ref{Solutionu22}).  Comparison with the numerical results shows that the
positive terms in the first square brackets in Eqs.~(\ref{Solutionu11}) and
(\ref{Solutionu22}) significantly reduce the gap between the analytical results
based on the FvK equations and the simulation data. 

Now, we resort to the geometric concepts of metric and curvature to analyze the
deformed lattice of the L-J particles in mechanical equilibrium for fully understanding
the inhomogeneity phenomenon.  A key observation is that the distribution of the
bond angle is sharply concentrated at $\pi/3$ for varying values of
$\gamma$.~\cite{SI} The deformation of the lattice is therefore approximately
angle-preserved.~\cite{rothen1993conformal, pieranski1989gravity} In
mathematics, strictly angle-preserved deformation is known as conformal
transformation.  Here, the quasi-conformal deformation of the lattice allows us
to approximately construct the metric of the following form on the undeformed
planar disk:
\begin{eqnarray}
ds^2 = \Lambda(\rho)^2 (d\rho^2 + \rho^2 d \varphi^2),\label{ds}
\end{eqnarray} 
where $\Lambda = d\ell/d\ell'$. $d\ell$ and $d\ell'$ are the length of the line element
at $\rho$ before and after the deformation, respectively. Equation~\ref{ds}
represents a purely geometric approach to understanding the deformation of the lattice. 

According to the Gauss's Theorema Egregium, the Gaussian curvature is solely
determined by the metric.~\cite{struik88a}  As determined by Eq.~(\ref{ds}), the
Gaussian curvature $K_\ell$ associated with the intrinsic inhomogeneity of the
lattice (regardless of its shape in three-dimensional Euclidean space)
is:~\cite{mughal2007topological,soni2018emergent} 
\begin{align}
\label{EqKEta} K_{\ell}=-\frac{1}{2}\Delta \ln \frac{\eta(\rho)}{\eta_0},
\end{align}
where the ratio of the densities $\eta(\rho)/\eta_0 = \Lambda(\rho)^{-2}$.
$\eta_0$ is the density of the undeformed lattice. $\Delta$ is the Laplacian
operator. The subscript in $K_{\ell}$ is to indicate that this curvature is
associated with the lattice; the Gaussian curvature of the spherical substrate
is denoted as $K_s$. From the geometric perspective, by stretching the lattice,
the spherical substrate essentially induces a Gaussian curvature in the
resulting inhomogeneous lattice.
Could the induced Gaussian curvature $K_\ell$, as governed by the
mechanical law, fully screen the curvature of the substrate geometry? In
other words, is $K_\ell=-K_s$?

To address this question, we further analyze the 
simulation data to address inquiry into the nature of the
induced curvature $K_{\ell}$. The plot of the integrated curvatures is presented
in Fig.~2(a). First of all, we see that the sign of $K_{\ell}$ is negative over
the entire lattice at varying degrees of stretching.  But the magnitude of the
integrated $K_{\ell}$ is always smaller than the integrated $K_{s}$. In other
words, the curvature of the spherical substrate could not be fully screened by
the induced curvature. From Fig.~2(b), we see that the ratio of the excess
integrated curvature and the integrated $K_{s}$ is rather stable. It implies
that the inhomogeneity created in the deformed lattice is to uniformize the
screening of the substrate curvature, although this screening is not
complete.

To explore the physical origin of the partial, but uniform screening phenomenon,
we first establish the relation between $K_{\ell}$ and the strain
tensor under the small deformation approximation:~\cite{SI}
\begin{eqnarray} \label{Kuij}
K_{\ell} = \frac{1}{2} \left[ \frac{1}{\rho}\frac{d}{d\rho}(u_{\rho\rho} + u_{\varphi\varphi}) +
\frac{d^2}{d\rho^2}(u_{\rho\rho} + u_{\varphi\varphi})  \right],
\end{eqnarray} 
Note that Eq.~(\ref{Kuij}) could be used to analyze the intrinsic curvature structure
in a series of problems related to the wrinkling of circular sheets under
tension or in differential growth.~\cite{cerda2003geometry, sharon2007geometrically} Inserting
Eqs.~(\ref{Solutionu11}) and (\ref{Solutionu22}) into Eq.~(\ref{Kuij}), we have 
\begin{align}\label{Eq-ratioKG}
\frac{K_s+K_{\ell}}{K_s} = &
\frac{1+\nu}{2}+O\left(\frac{\rho_0^2}{R^2}\right).
\end{align} 
Equation~(\ref{Eq-ratioKG}) clearly shows that $K_\ell\neq -K_s$ in general.
$K_\ell$ is determined by both the curvature of the substrate and microscopic
interaction; the later effect is reflected in the quantity of the Poisson's
ratio according to Eq.~(\ref{Eq-ratioKG}). Here, it is of interest to note
that the amount of excess curvature is completely determined by the Poisson's
ratio, and it is independent of other elastic moduli of the material. For our
elastic cap system, where $\nu_{\text{eff}}=0.31$, the relative excess curvature is about $0.65$ by
Eq.~(\ref{Eq-ratioKG}). This result agrees well with the simulation data in
Fig.~2.

Here, we shall emphasize that the screening mechanism by creating 
inhomogeneity with quasi-conformal ordering occurs in the elastic 
regime, and it is fundamentally different from the conventional screening 
scenario of plastic deformation based on topological defects.~\cite{Nelson1987,
	bowick2000interacting} In the plastic deformation of geometrically 
frustrated 2D crystals on curved space, topological defects tend to proliferate 
to screen the Gaussian curvature of the substrate surface to minimize the 
elastic free energy.~\cite{Nelson1987, bowick2000interacting} In our
system, the proliferation of topological defects is suppressed under the fixed
boundary condition. Our previous study on the frustration of L-J crystal
clusters on the sphere shows that, under the stress-free boundary condition, the
appearance of interior dislocations accompanies the formation of step structures
along the cluster contour.~\cite{yao2017topological} The fixed boundary
condition forbids the relative displacement of the boundary particles, and thus
suppresses the proliferation of interior topological defects.  As such, the
crystalline cap system deactivates the screening mechanism based on topological
defects, and adopts the fashion of creating inhomogeneity to fit the curved substrate.

We proceed to explore the plastic deformation regime, where the highly stretched
crystalline cap ultimately yields under the accumulated stress. Simulations
show that the cap is fractured as the value of $\gamma$ exceeds $7.8\%$,
regardless of the relative position of the entire lattice with respect to the
center of the cap. The global rupture of the cap is triggered by slightly
increasing $\gamma$ at the rate of $\delta \gamma$; $\delta\gamma$
is at the order of $10^{-6}$. A typical fracture pattern is
shown in Fig.~3(a); fracture patterns with three branches are also
observed.~\cite{SI} It is uniformly observed that the fracture takes a $90^{o}$
turn near the boundary. The radial and azimuthal fractures occur 
simultaneously. In this turning, the microscopic morphology of the
fracture transforms from the zigzag to the flat type, as illustrated in
Fig.~3(a).

To understand the formation of the featured rupture mode in Fig.3(a), we first
analyze the strain field in the central region. This region is isotropically
stretched, and it is subject to maximum stretching according to the solved
strain field.  In an isotopically stretched two-dimensional crystal, which type
of fracture will cost less energy? To address this question, we perform
numerical experiment with a planar L-J lattice under isotropic stretching whose
lattice spacing is $a$.  By comparing the energies of creating zigzag and flat
fractures of separation $\delta$, we obtain the phase diagram in Fig.3(b). It
turns out that the zigzag fracture is relatively energetically favored. This
explains the observed zigzag fracture in the central region of the cap.

The turning of the fracture is driven by the highly anisotropic strain field
near the boundary of the cap, as shown in the $u_{\rho\rho}$ and
$u_{\varphi\varphi}$ curves in Fig.1(c). Under the much larger radial strain
$u_{\rho\rho}$, the lattice tends to break along the azimuthal direction. To
conclude, the microscopic crystalline structure determines the morphology of the
central fracture pattern, and the anisotropy of the strain field is responsible
for the turning of the fracture near the boundary. The remarkable coexistence
of the radial and azimuthal fractures, which is absent in planar systems, is
essentially originated from the confluence of elasticity, curvature and the
microscopic crystalline structure. Note that curvature-driven fracture phenomena
are also reported in the deformation of nanoparticle monolayers
~\cite{mitchell2018conforming} and the 2D crystallization 
on the sphere. ~\cite{ortellado2022two}

Now, we discuss the energetics in the fracture of the crystalline cap.
Simulations show that the rupture event fundamentally changes the energy
landscape. The redistribution of the energy among individual particles is
summarized in Fig.~3(c). We see that the original single-peak profile splits into
multiple energy bands. Some particles jump to a few higher energy bands, 
but the energy of most particles is lowered. The total energy of the system is
significantly reduced by $8\%$. We also notice that the width of the original
energy profile shrinks in this process, indicating a reduced discrepancy in the
energy of the particles occupying the same energy band. 

\begin{figure}[] \subfigure[]{
		\includegraphics[height=0.20\textwidth]{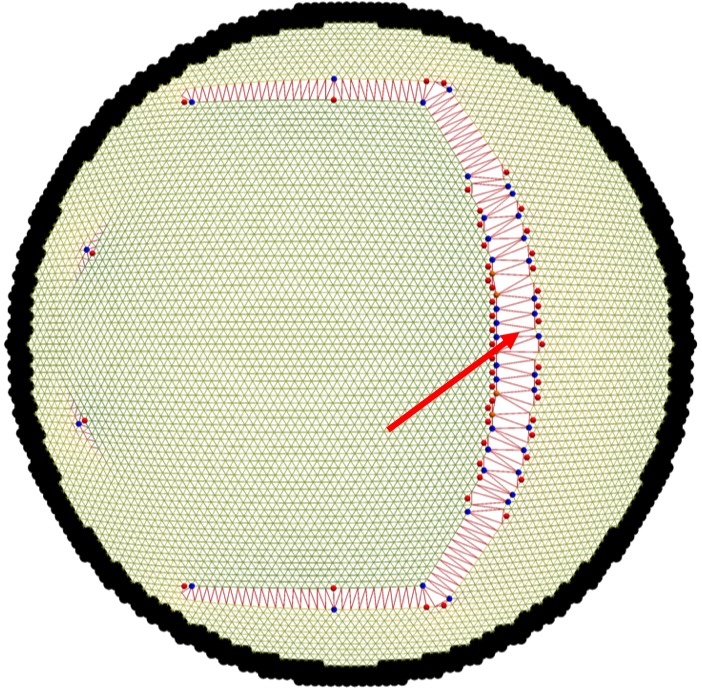}} 
	\subfigure[]{
		\includegraphics[width=0.25\textwidth]{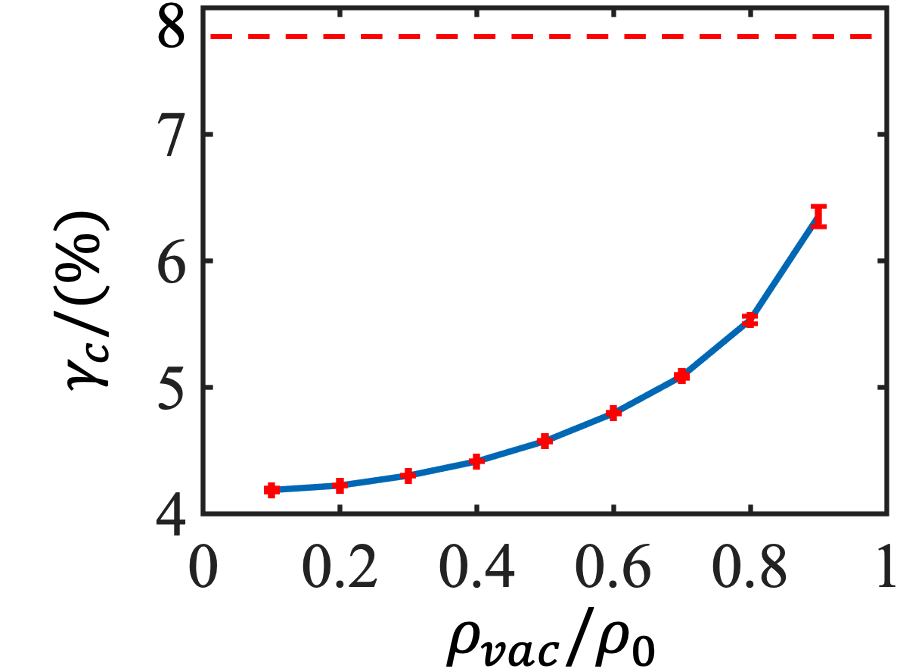}} 
	\caption{Controlling the fracture pattern and plastic deformation by exploiting the
		stress-concentration effect of the vacancy. (a) The presence of
		the vacancy, whose location is indicated by the arrow, leads to a fracture pattern
		that is distinct from the case of the vacancy-free cap in Fig.~3a. The red and blue
		dots represent five- and seven-fold disclinations. $\gamma=4.6\%$.
		$R=101.4r_m$. (b) Plot of
		the critical value for $\gamma$ versus the location of the vacancy. The dashed
		red line indicates the value of $\gamma_c=7.8\%$ for the case of the vacancy-free cap.
		$\rho_0=50r_m$.  The error bars are obtained by statistical analysis of the cases
		where the relative position of the entire lattice with respect to the center of the
		cap is randomly specified. }
	\label{Fig4} \end{figure}

The stress-steered rupture of the cap inspires us to explore the notion of
actively controlling the fracture by engineering the stress over the surface.
Specifically, we propose the strategy of introducing microscopic vacancies and
exploiting the stress-concentration effect to redistribute the stress and guide
the formation of the fracture.~\cite{timoshenko1951theory,yao2020stress}

A vacancy is created by removing a single particle in the crystalline cap. In
the presence of the vacancy, whose location is indicated by the arrow in
Fig.~4(a), the resulting fracture pattern is completely different from the case
in Fig.3(a).
The formation of the new pattern is attributed to the vacancy-driven 
amplified discrepancy of the radial and azimuthal strains surrounding the 
vacancy. The dominant radial strain ultimately leads to the azimuthal 
fracture passing through the vacancy.~\cite{timoshenko1951theory}
Instabilities caused by the stress-focusing effect of vacancies 
are also found in particulate elastic systems.~\cite{yao2020stress}
Simulations further show that introducing a
vacancy, especially near the center of the cap, could significantly reduce the
strength of the cap, as shown in Fig.4(b). The dashed line in Fig.4(b) indicates
the value of $\gamma_c$ for the case of vacancy-free cap. It is remarkable that
the microscopic operation of removing a single particle leads to macroscopic
effects in both aspects of fracture pattern and material strength.  More
examples of fracture patterns as controlled by the number and locations of
vacancies are provided in Supplemental Material.~\cite{SI}

\section{Conclusions}
In summary, we have used the crystalline cap model to reveal the generic
behaviors in the adaptivity of the two-dimensional crystalline systems on curved
space in both elastic and plastic regimes.  In the elastic regime, the geometric
concepts of metric and curvature provide a vigorous formalism for
understanding the observed inhomogeneous organization of the particles in the
quasi-conformal lattice. In the plastic regime, we observe the microscopic yield
of highly stressed caps, rationalize the emergence of the fracture patterns, and
demonstrate the strategy of exploiting the stress-concentration effect of
vacancies to actively control the fractures. This work shows the essential role
of geometry to foster insights into the deep connection between inhomogeneity,
stress and curvature. 

%\chen{\section*{Conflicts of interest}
%The authors declare no conflict of interest.}

\section*{Acknowledgements}
This work was supported by the National Natural Science Foundation of China 
(Grants No. BC4190050). The authors acknowledge the support from the Student 
Innovation Center at Shanghai Jiao Tong University.

%End of Main Text

%\bibliography{/Users/jingyuan/Research/Reference/Stress/Bib/thesis_v3.bib}
%\bibliography{ref}


\end{document}